\documentclass[aps,prl,reprint,preprintnumbers,superscriptaddress,amsmath,amssymb,bibnotes,longbibliography]{revtex4-2}
\usepackage{dcolumn}
\usepackage{bm}
\usepackage{graphicx}
\usepackage{amssymb}
\usepackage{amsmath}
\usepackage{epstopdf}
\usepackage{booktabs}
\usepackage{multirow}
\usepackage{siunitx}

\usepackage[pdfborder=001,colorlinks,linkcolor=blue,anchorcolor=blue,citecolor=blue,urlcolor=blue,filecolor=blue,menucolor=blue,runcolor=blue]{hyperref}
\newcommand{\tcr}{\textcolor{black}}

\begin{document}
\title{Semimetallic Kondo lattice behavior in YbPdAs with a distorted kagome structure}
\author{W. Xie}
\affiliation{Center for Correlated Matter and School of Physics, Zhejiang University, Hangzhou 310058, China}
\affiliation{Deutsches Elektronen-Synchrotron (DESY), Notkestrasse 85, 22607 Hamburg, Germany}
\author{F. Du} 
\affiliation{Center for Correlated Matter and School of Physics, Zhejiang University, Hangzhou 310058, China}
\author{X. Y. Zheng}
\affiliation{Center for Correlated Matter and School of Physics, Zhejiang University, Hangzhou 310058, China}
\author{H. Su}
\affiliation{Center for Correlated Matter and School of Physics, Zhejiang University, Hangzhou 310058, China}
\author{Z. Y. Nie}
\affiliation{Center for Correlated Matter and School of Physics, Zhejiang University, Hangzhou 310058, China}
\author{B. Q. Liu}
\affiliation{Key Laboratory of Neutron Physics and Institute of Nuclear Physics and Chemistry, CAEP, Mianyang 621900, PR China}
\author{Y. H. Xia}
\affiliation{Key Laboratory of Neutron Physics and Institute of Nuclear Physics and Chemistry, CAEP, Mianyang 621900, PR China}
\author{T. Shang}
\affiliation{Key Laboratory of Polar Materials and Devices (MOE), School of Physics and Electronic Science, East China Normal University, Shanghai 200241, People's Republic of China}
\author{C. Cao}
\affiliation{Center for Correlated Matter and School of Physics, Zhejiang University, Hangzhou 310058, China}
\author{M. Smidman}
\email{msmidman@zju.edu.cn}
\affiliation{Center for Correlated Matter and School of Physics, Zhejiang University, Hangzhou 310058, China}
\author{T. Takabatake}
\affiliation{Center for Correlated Matter and School of Physics, Zhejiang University, Hangzhou 310058, China}
\affiliation{Department of Quantum Matter, AdSE, Hiroshima University, Higashi-Hiroshima 739-8530, Japan}
\author{H. Q. Yuan}
\email{hqyuan@zju.edu.cn}
\affiliation{Center for Correlated Matter and School of Physics, Zhejiang University, Hangzhou 310058, China}
\affiliation{Zhejiang Province Key Laboratory of Quantum Technology and Device,
School of Physics, Zhejiang University, Hangzhou 310058, China}
\affiliation{State Key Laboratory of Silicon Materials, Zhejiang University, Hangzhou 310058, China}
\affiliation{Collaborative Innovation Center of Advanced Microstructures, Nanjing 210093, China}
\date{\today}

\begin{abstract}
We have synthesized YbPdAs with the hexagonal ZrNiAl-type structure, in which the Yb-atoms form a distorted kagome sublattice in the hexagonal basal plane. Magnetic, transport, and thermodynamic measurements indicate that YbPdAs is a low-carrier Kondo lattice compound with an antiferromagnetic transition at $T_\mathrm{N}$ = 6.6 K, which is slightly suppressed in applied magnetic fields up to 9 T. The magnetic entropy at $T_\mathrm{N}$ recovers only 33\% of $R\ln{2}$, the full entropy of  the ground state doublet of the Yb-ions. The resistivity displays a $-\ln T$ dependence between 30 and 15 K, followed by a broad maximum at $T\rm_{coh}$ = 12 K upon cooling. Below $T\rm_{coh}$, the magnetoresistance changes from negative to positive, suggesting a crossover from single-ion Kondo scattering processes at intermediate temperatures to coherent Kondo lattice behaviors at low temperatures. Both the Hall resistivity measurements and band structure calculations indicate a relatively low carrier concentration in YbPdAs. Our results suggest that YbPdAs could provide an opportunity for examining the interplay of Kondo physics and magnetic frustration in low carrier systems.

\begin{description}
\item[PACS number(s)]

\end{description}
\end{abstract}

\maketitle

\section{I. Introduction}

Magnetic frustration is an important means for tuning \tcr{heavy-fermion systems}, as exemplified by the proposed global phase diagram \cite{vojta1,vojta2,Si2010,coleman2010}, which considers both frustration and Kondo hybridization as tuning parameters. Upon increasing the frustration strength within the weak Kondo hybridization regime, magnetic order can be destroyed yielding a spin-liquid ground state with localized 4\textit{f} electrons. With a stronger Kondo hybridization, the system may directly change from a spin liquid to a heavy Fermi-liquid state, which could correspond to either a quantum phase transition or a crossover \cite{Si2010, coleman2010, Si2014}. 

 \tcr{Rare-earth compounds with the hexagonal ZrNiAl-type structure (space group \emph{P}$\bar{6}2$\emph{m}, No.189),  where magnetic rare-earth atoms form a distorted kagome sublattice, exhibit a range of frustration induced behaviors such as partial ordering \cite{Donni1996}, non-collinear magnetism \cite{TmAgGe} and spin ice states \cite{HoAgGe}. Some Ce- and Yb- based compounds with this structure exhibit remarkable manifestations of the interplay between magnetic frustration and the Kondo effect  \cite{RTX},} where systematic investigations have revealed unusual physical properties, such as a possible extended spin liquid phase in CePdAl \cite{CePdAl1,CePdAl2,CePdAl2016,CePdAl2019,CePdAl4,CePdAl2022}, as well as complex phase diagrams together with novel quantum criticality in both CePdAl and YbAgGe \cite{CePdAl1,CePdAl2,CePdAl2016,CePdAl2019,CePdAl4, CePdAl2022, YbAgGe1,YbAgGe3,YbAgGe2010}. 
However, evidence for the coexistence of the Kondo effect and significant magnetic frustration has only been found in a limited number of these materials \cite{CePdAl4,YbAgGe3, CeRhSn, CeIrSn, YbPtIn2000}, and frustration induced novel phenomena have not been reported for many isostructural compounds \cite{RTX, YbRhSn2,YbPtSn1,YbPdSn1}. This poses the question as to conditions necessary for realizing strong magnetic frustration in such Kondo intermetallics.  

\begin{figure}[!htb]
     \begin{center}
     \includegraphics[width=\columnwidth,keepaspectratio]{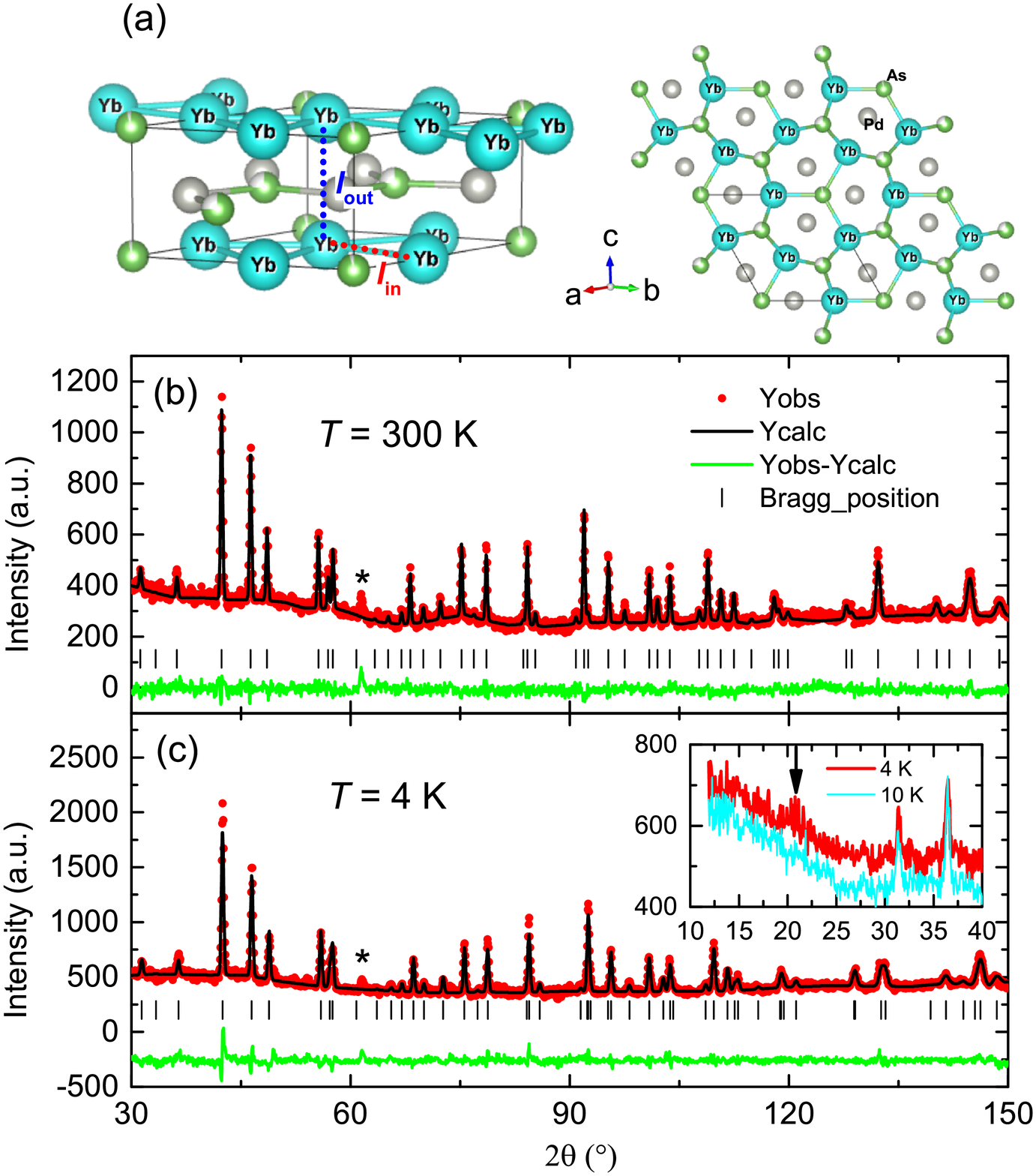}
     \end{center}
     \caption{(Color online) (a) Crystal structure of YbPdAs, where the right image highlights the distorted kagome lattice formed by Yb atoms in the hexagonal basal plane. Powder neutron diffraction patterns of YbPdAs are shown at (b) 300 K, and (c) 4 K. The solid black lines show the calculated patterns from the Rietveld refinements. The inset of (c) shows the data at 4 K (below $T_\mathrm{N}$) and 10 K (right above $T_\mathrm{N}$) at low angles, where the 10 K data has been shifted downwards for comparison.  \tcr{
     The asterisk and arrow indicate a peak from the impurity phase Yb$_2$O$_3$, and a possible additional peak present only at 4 K, respectively.}}
     \label{fig1}
\end{figure}

To examine this issue, the nearest neighbor (NN) distances between the Ce/Yb atoms, both within ($l_{in}$) and out of ($l_{out}$) the basal plane (Fig.~\ref{fig1} (a)), have been summarized in Table. \ref{table1} for several compounds, together with  some of their physical properties. The difference between the in and out of plane distances $\Delta l$ = $l_{out} - l_{in}$ is also calculated, shown together with $\Delta l$/$l_{in}$, which potentially serves as an approximate measure of the effective dimensionality of the magnetic sublattice. It can be seen that for CePdAl and YbAgGe, which exhibit frustration-induced novel phenomena, both $\Delta l$ and $\Delta l$/$l_{in}$ are considerably larger, indicating that the strong frustration in these systems may in-part be a consequence of the reduced dimensionality. Such a reduced dimensionality has been evidenced by their highly anisotropic transport and magnetic properties \cite{YbAgGe1, YbAgGe2010,CePdAl1}. For YbPdAs, $\Delta l$ = 0.22 ${\AA}$ and $\Delta l$/$l_{in}$ = 6.0\% \cite{YbPdAs-structure}, which while smaller than the aforementioned CePdAl/YbAgGe, is relatively large compared to several others, yet other physical properties are not yet reported.

Here we have synthesized polycrystalline YbPdAs and characterized its physical properties using transport, magnetic and thermodynamic measurements. Our results indicate that YbPdAs is a Kondo lattice compound with a second-order magnetic phase transition at $T_\mathrm{N}$ = 6.6 K. In-field measurements show that $T_\mathrm{N}$ is rather robust to applied magnetic field up to 9 T. Band structure calculations show a small Fermi surface with the presence of both electron and hole carriers, indicating a low carrier nature, in line with the relatively large resistivity and Hall coefficient.
Our studies suggest that YbPdAs is a low carrier Kondo system where magnetism may potentially coexist with magnetic frustration.

\begin{table*}[htb]
    \centering
    \caption{A summary of NN distances between Ce/Yb atoms in several compounds with the ZrNiAl-type structure, where $l_{in}$, $l_{out}$, and $\Delta l$ = $l_{out} - l_{in}$ are NN distances in the hexagonal plane, out of the plane, and their differences, respectively. }
    \setlength{\tabcolsep}{2mm}
    \begin{tabular}{lccccc}
    \toprule
      Compounds & $l_{in}$ (${\AA}$) & $l_{out}$(${\AA}$) & $\Delta l$(${\AA}$) & $\Delta l$/$l_{in}$(\%)  & Properties \\
      \midrule
     CePdAl \cite{CePdAl1,CePdAl2016,CePdAl2019, CePdAl-structure}  & 3.79 & 4.23 & 0.44 & 11.6 & \begin{tabular}[c]{@{}c@{}}
           Frustration  \\  Quantum criticality \\ (under pressure/magnetic field)
     \end{tabular}    \\
    \hline
     YbAgGe \cite{YbAgGe1, YbAgGe3}  & 3.70 & 4.14&	0.44 & 11.9 & 
      \begin{tabular}[c]{@{}c@{}}
           Frustration  \\  Quantum criticality \\ (under magnetic field)
     \end{tabular}    \\
     \hline
      YbPdAs \cite{YbPdAs-structure}  & 3.69 & 3.91 & 0.22&  6.0 &	?  \\
     \hline
     YbPtSn \cite{YbPtSn1}  & 3.87  & 3.93  & 0.06 & 1.6& $T_\mathrm{N}$ = 3.3 K \\
     \hline
    CePdIn \cite{CePdIn}   & 4.04 & 4.08  & 0.04&1.0 &	$T_\mathrm{N}$ = 1.6 K  \\
     \hline
     YbPtIn \cite{YbPtIn1,YbPtIn3}  & 3.79 &3.77  & --0.02 & --0.5 &  \begin{tabular}[c]{@{}c@{}}
           Weak frustration  \\  $T_\mathrm{N1}$ = 3.4 K\\$T_\mathrm{N2}$ = 1.4 K
     \end{tabular}    \\	  
     \hline
     $\alpha$-YbPdSn \cite{YbPdSn1} &3.98&	3.76  &--0.22& --5.5&	$T_\mathrm{N}$ = 0.25 K  \\
     \hline
     YbRhSn \cite{YbRhSn2}   &3.96& 3.67  &--0.29 & --7.3& \begin{tabular}[c]{@{}c@{}}
             $T_\mathrm{N1}$ = 1.85 K\\$T_\mathrm{N2}$ = 1.4 K
     \end{tabular}    \\	 
    \bottomrule
    \end{tabular}
    \label{table1}
\end{table*}

\section{II. Experimental methods}
Polycrystalline YbPdAs and LuPdAs (as a non-magnetic reference) were synthesized using a solid state reaction method. The binary material PdAs$_2$ was first synthesized by reacting stoichiometric Pd powder (99.9\%, Alfa Aesar) and As granules (99.999\%, Alfa Aesar) at 700$^\circ$C for two days. The PdAs$_2$ powder was then mixed with Yb or Lu pieces (99.9\%, Alfa Aesar) and Pd powder. The mixture was pelletized, wrapped in Ta-foil and sealed in an evacuated quartz ampule. The quartz ampule was slowly heated up to 900$^\circ$C, kept there for three days before being furnace-cooled down to room temperature.
The formation of the much more stable YbAs phase appears to hinder the growth of YbPdAs single crystals \cite{YbAs}.

The crystal structure was characterized at room temperature by powder x-ray diffraction on a PANalytical X'Pert MRD diffractometer with Cu-K$\alpha$ radiation monochromated by graphite. Further characterization using neutron diffraction at 300, 200, 10 and 4 K was performed using the high-resolution powder diffractometor Xuanwu at the China Mianyang Research Reactor with a neutron wavelength of 1.8846 ${\AA}$. Rietveld refinements were performed on the obtained patterns using the {\sc fullprof} package.

The electrical resistivity and heat capacity were
measured using a Quantum Design Physical Property Measurement System (PPMS-9T) equipped with a $^3$He insert.
The magnetization was measured down to 2 K using a Quantum Design Magnetic Property Measurement System 5T,
and a vibrating sample magnetometer (VSM) on a PPMS. 
 Hall measurements were performed using a PPMS-9T, utilizing a four-wire-method on a well-polished sample, where the transverse contribution ($\rho_{xx}$) was removed by subtracting the negative field data from those at corresponding positive fields.

Density  functional  theory  (DFT)  calculations were performed using the plane-wave projected augmented wave method as implemented in the Vienna \emph{ab initio} simulation package ($\texttt{VASP}$) code \cite{vasp1,vasp2}. A  plane-wave  basis  up  to  400  eV  and  6 $\times$ 6 $\times$ 12 $\Gamma$-centered $K$ mesh were used to integrate over the Brillouin zone. The 4\textit{f} electrons are assumed to be core states in the calculations and spin-orbit coupling (SOC) was considered in all calculations. The band structures from VASP were fitted to tight-binding Hamiltonians using the maximally projected Wannier function method \cite{vasp3}. The Fermi surfaces are then obtained by interpolating the band structure to a 100 $\times$ 100 $\times$ 100 dense $K$ mesh using this tight-binding Hamiltonian.

\begin{figure}[ht]
\begin{center}
     \includegraphics[width=\columnwidth,keepaspectratio]{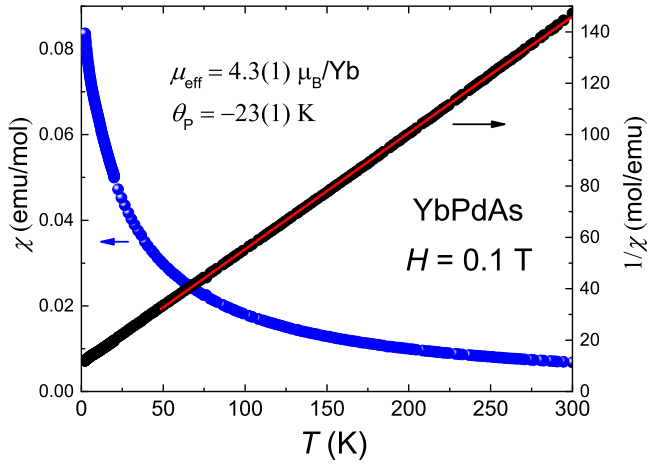}
     \end{center}
     \caption{(Color online) Temperature dependence of the magnetic susceptibility  $\chi$(\emph{T}), and inverse magnetic susceptibility 1/$\chi(T)$ from 300 to 2 K measured in an applied field of 0.1 T. The fit using the Curie-Weiss law to the high-temperature data of 1/$\chi(T)$ is shown by a red solid line.}
\label{fig2}
\end{figure}

\section{III. Results and discussion}
\subsection{A. Zero field characterization}

\begin{table*}[ht]
    \centering
    \caption{The lattice parameters and atomic positions of YbPdAs obtained from Rietveld refinements on the powder neutron diffraction data at 4 and 300K, together with  the profile factor $R_p$ and the weighted profile factor $R_{wp}$. }
     \setlength{\tabcolsep}{1.8mm}
    \begin{tabular}{c|c|ccc|ccc}
    \toprule
      & & \multicolumn{3}{c}{4 K} & \multicolumn{3}{c}{300 K}  \\
         \midrule
       \multirow{2}*{Lattice parameters} & & $a$(${\AA}$) &  $b$(${\AA}$) & $c$(${\AA}$) & $a$(${\AA}$) & $b$(${\AA}$) & $c$(${\AA}$) \\
        &~ & 6.963(1) & 6.963(1) & 3.917(1) & 7.002(1) & 7.002(1) & 3.914(1) \\
      \midrule
      \multirow{5}*{Atomic coordinates}& & $x$ & $y$ & $z$ & $x$ & $y$ & $z$ \\
      & Yb &  0.5979(5) & 0.0000 & 0.0000 & 0.5967(6) & 0.0000 & 0.0000\\
      & Pd &   0.2673(9) & 0.0000 & 0.5000 &  0.2618(13) & 0.0000 & 0.5000 \\
      & As$_1$ &   0.3333 &  0.6667 & 0.5000 & 0.3333 &  0.6667 & 0.5000 \\
      & As$_2$ &  0.0000 &  0.0000 &  0.0000 &  0.0000 & 0.0000 &  0.0000\\
     \midrule
      \multirow{2}*{R-factors} & &$R_{p} (\%)$ & $R_{wp}(\%)$  &  & $R_{p}(\%)$ & $R_{wp}(\%)$  & \\
     & &  3.75  &    4.97    &         &    3.16  & 4.06 &  \\
    \bottomrule
    \end{tabular}
    \label{table3}
\end{table*}

Figure~\ref{fig1} shows powder neutron diffraction patterns for YbPdAs at 300 and 4 K, together with the simulated patterns from Rietveld structural refinements. A small unmatched peak at 2$\theta \approx$ 62.0$^\circ$ is due to a Yb$_2$O$_3$ impurity phase \cite{Yb2O3}. The refined crystal structure parameters are listed in Table. \ref{table3}. The obtained lattice parameters are $a$ = $b$ = 7.002(1) ${\AA}$ and $c$ = 3.914(1) ${\AA}$ at 300 K, being consistent with the previous report \cite{YbPdAs-structure}. Accordingly, values of $l_{in}$ and $l_{out}$ of 3.691(1) ${\AA}$ and 3.914(1) ${\AA}$ are obtained, respectively, at 300 K (3.678(1) ${\AA}$ and 3.917(1) ${\AA}$ at 4 K), yielding $\Delta l$ of 0.223(1) ${\AA}$ at 300 K  (0.239(1) ${\AA}$ at 4 K). This indicates a moderate two-dimensionality for the Yb-sublattice. Consistent neutron diffraction profiles at 4 K and 300 K exclude a possible structural transition upon cooling.   \tcr{
There are no clearly resolved additional peaks at 4 K compared to 10 K, but there is a weak bump at 2$\theta = 20.8^\circ$, as shown in the inset of Fig.~\ref{fig1}(c). The absence of clearly resolved
magnetic peaks at 4 K may be due to the small size of the ordered moments, which can be reduced by CEF effects, Kondo screening, and magnetic frustration. Further measurements on a high-flux cold neutron diffractometer at lower temperatures may be able to resolve magnetic Bragg peaks in YbPdAs corresponding to the magnetic order (see below). }

 \begin{figure}[!t]
\centering\includegraphics[width=0.95\columnwidth,keepaspectratio]{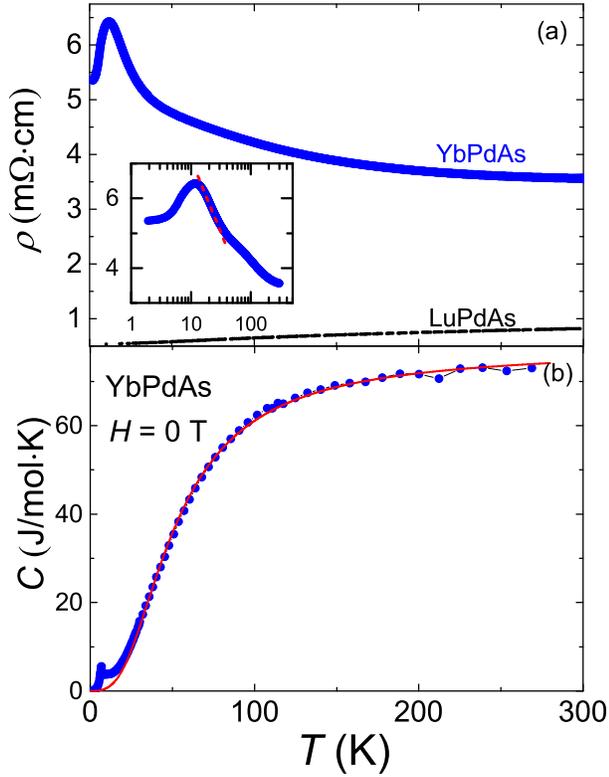}
\caption{(Color online) (a) Temperature dependence of the resistivity $\rho(T)$ of YbPdAs and LuPdAs from 300 to 1.9 K. \tcr{The inset shows the data of YbPdAs on a logarithmic temperature scale, where the dashed line corresponds to a logarithmic temperature dependence.}  (b) Temperature dependence of the heat capacity \emph{C}(\emph{T}) of YbPdAs, and the corresponding fit using the Debye model (solid red line).  }
\label{fig3}
\end{figure}

 Figure~\ref{fig2} shows the temperature dependence of the magnetic susceptibility $\chi(T)$ (and its inverse 1/$\chi(T)$) of YbPdAs from 300 to 2 K. 1/$\chi(T)$ for $T \geq $ 20 K exhibits a Curie-Weiss behavior $\chi = C/(T-\theta\rm_p)$, where \textit{C} and $\theta\rm_p$  are the Curie constant and Weiss temperature, respectively. The derived effective moment $\mu_{eff}$ of 4.3(1) $\mu_{\rm_B}$/Yb indicates a trivalent nature of the Yb ions, while the negative $\theta\rm_P$ of --23(1) K suggests dominant antiferromagnetic interactions between Yb moments. 
 
 In Fig.~\ref{fig3},  the temperature dependence of the electrical resistivity $\rho(T)$ and heat capacity $C(T)$ are shown. The $\rho(T)$ of YbPdAs increases with decreasing temperature down to 12 K, where there is a broad maximum. This is in contrast with the metallic behavior for  isostructural nonmagnetic LuPdAs. This implies that there is an important role played by hybridization between conduction electrons and 4\textit{f}-electrons ($c-f$ hybridization).
\tcr{As shown in the inset of Fig.~\ref{fig3}(a), $\rho(T)$ shows a logarithmic temperature dependence between 15--30 K, indicating the presence of the Kondo effect.} The resistivity is of order of m$\Omega$-cm for YbPdAs, which is at least one order of magnitude larger than that typical for Yb-based metallic Kondo compounds \cite{YbPtIn2000}. Such a large resistivity is similar to that for YbRh$_3$Si$_7$, where the negative slope of the resistivity is ascribed to Kondo scattering in the framework of a semimetallic electronic structure \cite{YbRh3Si7}.  As shown in the main panel of Fig.~\ref{fig3}(b), $C(T)$ at $T \geq $ 30 K can be well fitted by a Debye model (red solid line), with a Debye temperature $\Theta_\mathrm{D}$ of 210(5) K. This fitting is used to estimate the phonon contribution to $C(T)$ across the whole temperature range.

\begin{figure}[!bt]
\centering\includegraphics[width=\columnwidth,keepaspectratio]{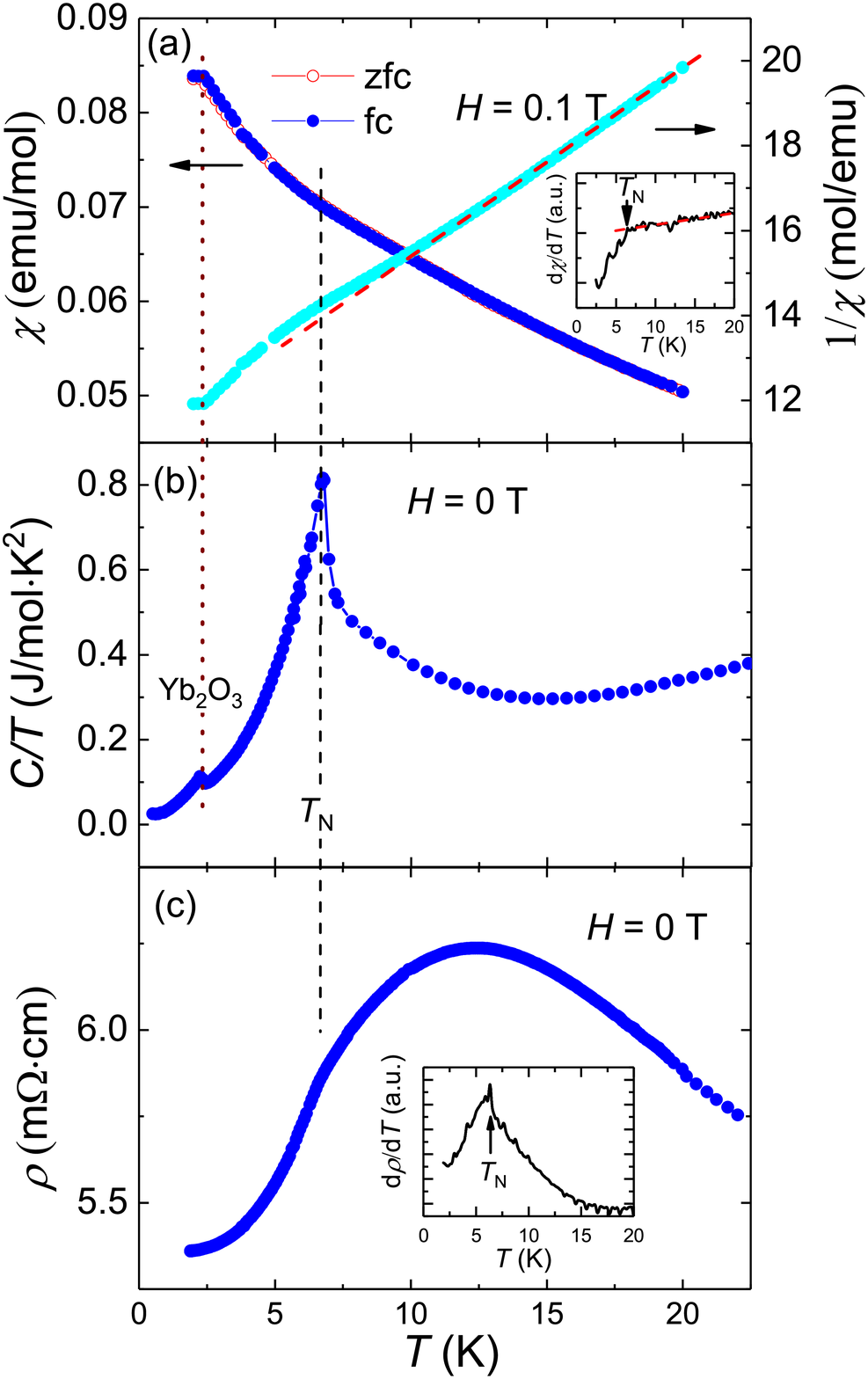}
\caption{(Color online)  Low temperature data of  (a) magnetic susceptibility $\chi$(\emph{T}), (b) specific heat $C(T)/T$, and (c) resistivity $\rho(T)$ of YbPdAs. \tcr{The red dashed line in (a) corresponds to the high temperature linear behavior of 1/$\chi$(\emph{T}).} The insets in (a) and (c) show the derivatives of  $\chi$(\emph{T}) and $\rho(T)$, respectively.}
\label{fig4}
\end{figure}

\begin{figure}[!htb]
  \centering
  \includegraphics[width=1\columnwidth,keepaspectratio]{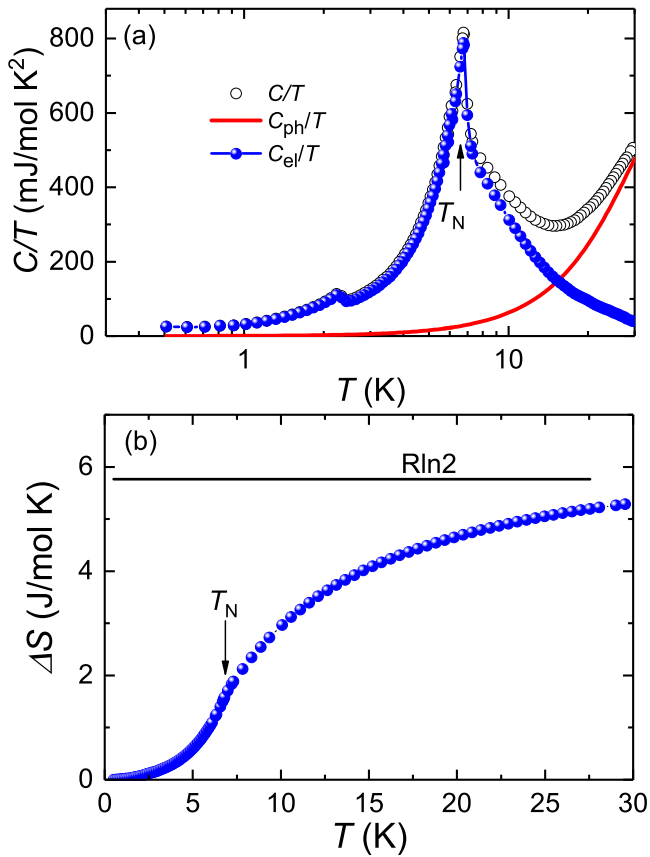}
  \caption{(Color online) (a) Specific heat $C(T)/T$ of YbPdAs shown on a logarithmic temperature scale (black circles), together with with the electronic contribution $C_{el}(T)/T$ (blue circles). (b) The  magnetic entropy $\Delta S$ obtained by integrating the  $C_{el}(T)/T$ data in (a). }
\label{fig5}
\end{figure}

The low temperature data of $\chi(T)$, $C(T)/T$ and $\rho(T)$ of YbPdAs are displayed in Fig.~\ref{fig4}. Here 1/$\chi(T)$ displays a small deviation from the Curie-Weiss law below about 12 K, where $C(T)/T$ starts to increase significantly, and $\rho(T)$ shows a broad maximum. These features are consistent with the gradual development of Kondo coherence upon cooling, where the peak temperature in $\rho(T)$ is regarded as the coherence temperature $T\rm_{coh}$. At around 6.6 K, $\chi(T)$ has a change of slope (see inset of Fig.~\ref{fig4}(a)), $C(T)/T$ exhibits a sharp $\lambda$-shaped peak, and there is an anomaly in $\rho(T)$. Furthermore, no discernible differences are found in $\chi(T)$ data between field-cooling and zero-field cooling, as shown in Fig.~\ref{fig4}(a). These data indicate a second order magnetic phase transition at $T_\mathrm{N}$ = 6.6 K, which is most likely antiferromagnetic (AFM). Note that the weak anomalies at 2.3 K are due to a small amount of antiferromagnetic Yb$_2$O$_3$, which can be suppressed by a small applied magnetic field, as reported for other Yb-based compounds containing Yb$_2$O$_3$ as an impurity \cite{YbPtIn2000}. As displayed in Fig.~\ref{fig4}(a), the transition at $T_\mathrm{N}$ leads to a relatively weak anomaly in $\chi(T)$ in YbPdAs. Such weak anomalies in $\chi(T)$ at $T_\mathrm{N}$ have been observed in a number of Yb-based antiferromagnets \cite{YbRh3Si7,YbIr3Si7_arXiv}, and the polycrystalline nature of the samples used here may further make the signature in $\chi(T)$ difficult to resolve. 

Figure~\ref{fig5}(a) shows $C(T)/T$ together with the electronic contribution \emph{C}$_{el}(T)/T$ obtained after subtracting the phonon contribution (red solid line in Fig.~\ref{fig3}). It can be seen that \emph{C}$_{el}(T)/T$ displays a $-\ln T$ dependence immediately above $T_\mathrm{N}$ (in the range of 7--15 K). This feature has been observed previously in the specific heat of CePdAl \cite{CePdAl4},  which was attributed to enhanced spin fluctuations arising from the underlying geometric frustration. For YbPdAs, the $-\ln T$ increase in $C(T)/T$ is concomitant with the development of a coherence peak in $\rho(T)$. \emph{C}$_{el}/T$ flattens below 1 K to 25 mJ/mol-K$^2$, indicating a small value of the Sommerfeld coefficient $\gamma$ in the AFM state. This could be related to the low carrier densities, which may lead to a shortage of conduction electrons in the $c-f$ hybridization process (so-called Kondo exhaustion \cite{kondo_exhaustion1,kondo_exhaustion2}), giving rise to a reduced $\gamma$ value \cite{CeNi2As2}. On the other hand, the reduced $\gamma$ is not likely due to the opening of an energy gap, given the decrease of $\rho(T)$ at low temperatures. 

 The relative magnetic entropy $\Delta S$ is calculated by integrating \emph{C}$_{el}/T$ from 0.4 K (the contribution below that is negligible), which is shown in Fig.~\ref{fig5}(b). The entropy released up to \emph{T}$_\mathrm{N}$ is only about 0.33$R\ln$2, which is much less than $R\ln$2, the expected entropy corresponding to the CEF ground state  doublet. The integrated entropy reaches $R\ln$2 at $T \geq$ 30 K. In YbPdAs, both the Kondo effect and frustration may lead to such a reduced entropy at the transition.

\begin{figure}[htb]
\centering
\includegraphics[width=0.95\columnwidth, keepaspectratio]{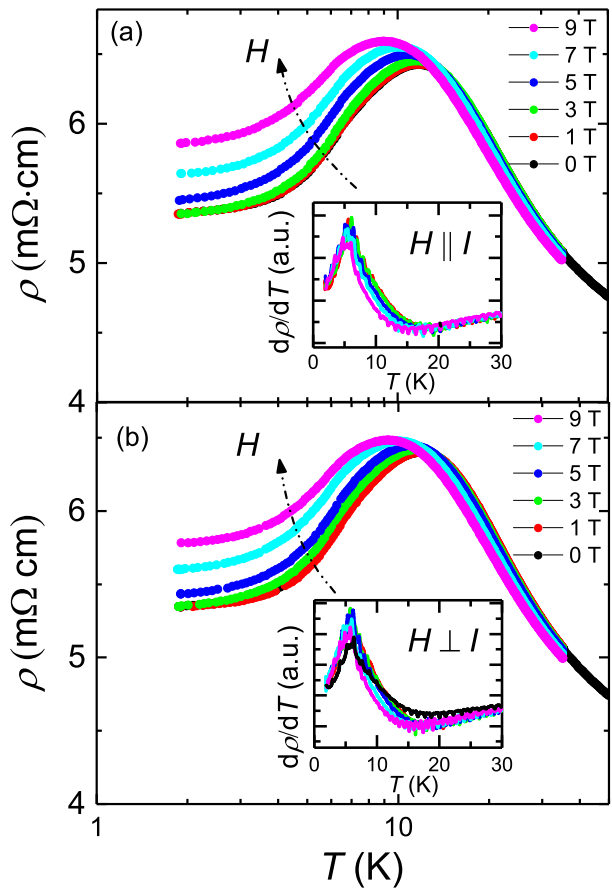}
\caption{(Color online) $\rho(T)$ of YbPdAs under magnetic fields up to 9 T in two different field-current configurations: (a) $H \parallel I$ and (b) $H \perp I$. The insets show the corresponding temperature derivatives.}
\label{fig6}
\end{figure}

\subsection{B. In-field properties}

\begin{figure}[ht]
\centering\includegraphics[width=1\columnwidth]{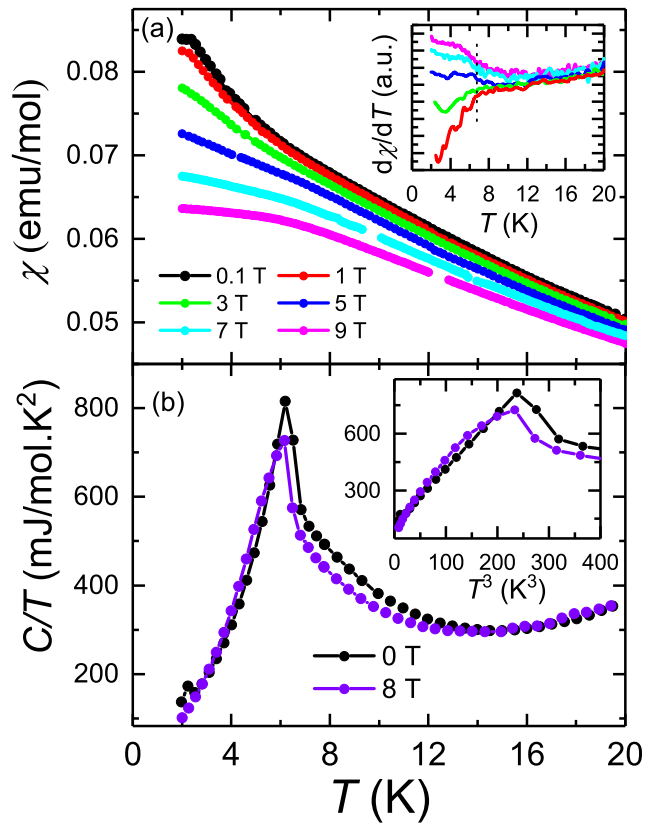}
\caption{(Color online) (a) Temperature dependence of the magnetic susceptibility $\chi$(\emph{T}), and (b) specific heat $C(T)/T$ measured in different applied magnetic fields. The inset of (a) shows the temperature dependence of $d\chi/dT$, while the inset of (b) shows the $T^3$ dependence of $C(T)/T$.}
\label{fig7}
\end{figure}

Figure~\ref{fig6} shows $\rho(T)$ under different  magnetic fields up to 9 T, applied parallel ($H \parallel I$) and perpendicular ($H \perp I$) to the current direction. 
For both field directions, the coherence peaks are suppressed to lower temperature with increasing magnetic field, while $\rho(T)$ at $T < T\rm_{coh}$ increases. These observations are not consistent with a single-ion Kondo impurity model, where $\rho$ at lower temperatures would be also suppressed by external magnetic fields \cite{UBe13_MR,YbPtSn1,YbPtIn2000}. The field dependence of $T_\mathrm{N}$ can be observed from the $d\rho(T)/dT$ shown in the insets, where there is only a very small shift of $T_\mathrm{N}$ to lower temperature with field. This is further corroborated by $\chi(T)$ and $C(T)$ measured under applied magnetic fields up to 9 T, shown in Fig.~\ref{fig7}, which show that $T_\mathrm{N}$ is almost field-independent. The rate at which $\chi(T)$ increases below $T_\mathrm{N}$ varies in different applied fields, as shown by the derivative $d\chi/dT$ (see inset of Fig.~\ref{fig7}(a)). Such behavior in $\chi (T)$ suggests that although $T_\mathrm{N}$ is robust to magnetic fields, the underlying magnetic structure and/or excitations gradually vary with fields, possibly due to canting of the spins. Furthermore, $C(T)/T$ in zero-field exhibits a $T^3$ behavior just below $T_\mathrm{N}$, but gradually deviates from this behavior with increasing field, as shown in the inset of Fig.~\ref{fig7}(b). The peak height in $C(T)$ at $T_\mathrm{N}$ is slightly suppressed at high field, while the $-\ln T$ behavior above $T_\mathrm{N}$ remains.

\begin{figure}[!ht]
\centering\includegraphics[width=\columnwidth,keepaspectratio]{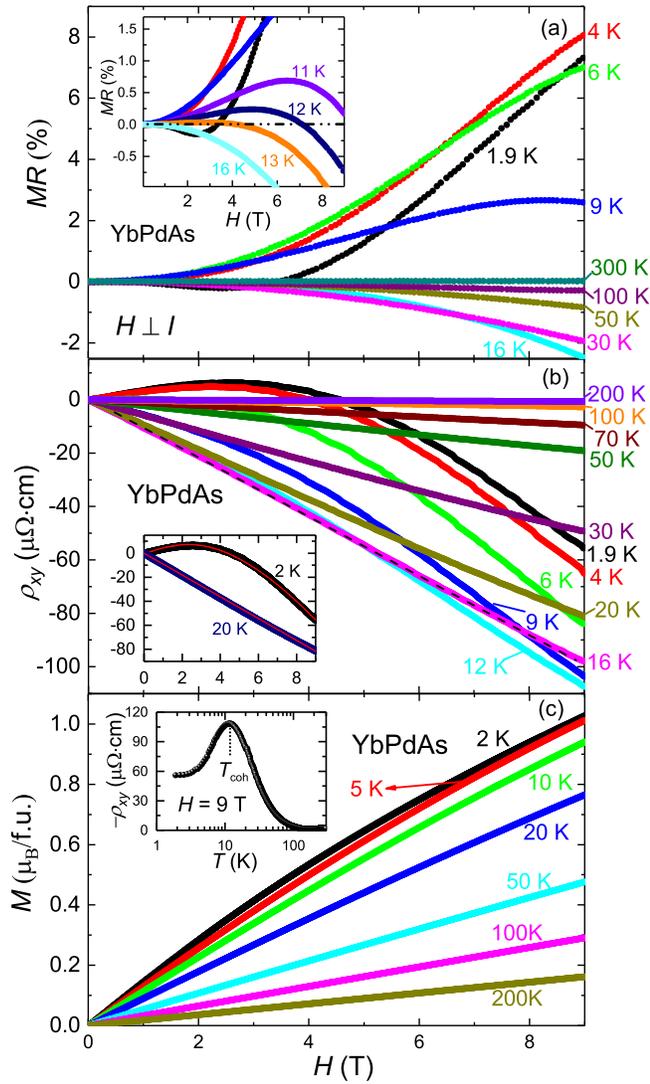}
\caption{(Color online)  (a) Field dependent magnetoresistance (MR) at different temperatures from 200 K to 1.9 K. The inset shows the enlarged data near $T\rm_{coh}$ = 12 K; (b) Field dependent Hall resistivity $\rho_{xy}$(\textit{H}) at different temperatures from 300 K to 1.9 K. The inset shows the fit of $\rho_{xy}$(\textit{H}) at 2 K and 20 K (see text); (c) Field dependent magnetization \textit{M}(\textit{H}) at different temperatures. The inset in (c) shows the temperature dependence of $-\rho_{xy}$(\textit{H}) at $H = 9$ T.}
\label{fig8}
\end{figure}

The magnetoresistance MR ( = $[\rho(H)-\rho(0)]/\rho(0) \times$ 100\% ) and Hall resistivity $\rho_{xy}$ at different temperatures are shown in Fig.~\ref{fig8}. There is a clear difference in the field dependent behavior for temperatures above and below $T_\mathrm{coh} \approx$ 12 K. For $T \geq T\rm_{coh}$, the MR is negative and $\rho_{xy}$ displays a linear field dependence; while for $T \leq T\rm_{coh}$, the MR starts to become positive (inset of Fig.~\ref{fig8}(a)), and $\rho_{xy}$ gradually becomes nonlinear. Considering that Kondo coherence develops below $T\rm_{coh}$, the different MR behaviors could be due to different quasiparticle scattering mechanisms. For $T \geq T\rm_{coh}$, incoherent scattering arising from single-ion Kondo processes may be suppressed by magnetic fields, leading to a negative MR, similar to many other Yb-based heavy fermion compounds \cite{YbPtIn2000, YbPtSn1}. However, below $T\rm_{coh}$, Kondo coherence greatly suppresses the scattering and therefore applied magnetic fields destroying the coherence will lead to an increased resistivity \cite{CeCu2Si2, CeAl3}. As shown in the inset of Fig.~\ref{fig8}(a), the MR at temperatures near $T\rm_{coh}$ increases first and then decreases, implying competition between the two mechanisms. Note that the weak negative MR at low fields at 1.9 K likely arises from the suppression of the AFM ordering of the Yb$_2$O$_3$ impurity.

The field dependence of $\rho_{xy}$ shows linear behavior at $T \geq T\rm_{coh}$ with a negative slope, indicating the dominance of electron type charge carriers. Meanwhile, the increasingly nonlinear $\rho_{xy}(H)$ upon cooling below $T\rm_{coh}$ is due to the presence of an anomalous Hall effect (AHE) arising from the enhancement of Kondo coupling strength \cite{RMP-AHE}. For a Kondo lattice compound, the dominant AHE mechanism is skew scattering where the conduction electrons are asymmetrically scattered by the magnetic impurity potentials, leading to a contribution that varies linearly with the electrical resistivity \cite{Smit1955,Smit1958,Fert1987, Nair2012}. The temperature dependence of the Hall resistivity $-\rho_{xy}(T)$ is shown in the inset of Fig.~\ref{fig8}(c). It can be seen that $-\rho_{xy}(T)$ shows similar behavior to the resistivity $\rho(T)$, with a broad hump centered at $T\rm_{coh}$.

The carrier density $n$ at different temperatures can be estimated using a single-band model, i.e., $R_\mathrm{H} = \rho_{xy}/B = 1/(ne)$. This yields a considerable decrease in $n$ from about $ 6.2\times10^{21}$ cm$^{-3}$ at 200 K to $5.24\times10^{19}$ cm$^{-3}$ at 12 K.
However, although linear $\rho_{xy}(H)$ behavior is observed for $T \geq$ 12 K, the Hall resistivity and MR data cannot be accounted for by a semi-classical two-band model \cite{two_band_model}, which especially cannot account for the negative MR. In YbPdAs, two band effects, magnetic scattering, and Kondo hybridization may all contribute to the transport behavior, which makes the different contributions difficult to disentangle. Note that the DFT calculations (see below) suggest that YbPdAs is a two band system with the presence of both electron and hole carriers.

The field dependent magnetization $M(H)$ are shown in Fig.~\ref{fig8}(c), which increase smoothly at all temperatures without saturation up to 9 T. The absence of spontaneous magnetization at 2 K corroborates the AFM nature of the magnetically ordered state, while the absence of a metamagnetic transition at 2 K is consistent with the robustness of $T_\mathrm{N}$ observed in $\chi(T)$ and $C(T)/T$. To quantitatively characterize the Hall resistivity, we have analyzed $\rho_{xy}(H)$ with \cite{CeNi2As2}:
\begin{equation}
    \rho_{xy} = R_HH + R_s \mu_0 M (H),
\label{eq1}
\end{equation}
where $R_\mathrm{H}$ is the normal Hall coefficient (assuming single-band behavior), and the second term characterizes the AHE contribution, \tcr{with $M(H)$ being the experimental data displayed in Fig.~\ref{fig8}(c).} As shown in the inset of Fig.~\ref{fig8}(b),  $\rho_{xy}(H)$ at 2 and 20 K are reasonably fitted by Eq.~\ref{eq1}, with  $R_\mathrm{H} = -46.1$ $\mu\Omega$cm/T, $R_s \mu_0$ = 350 $\mu\Omega$cm/($\mu_B$/f.u.) for 2 K, and $- 10.1$ $\mu\Omega$cm/T, 4 $\mu\Omega$cm/($\mu_B$/f.u.) respectively for 20 K. This corresponds to electron-type carriers with $n$ = 1.35 $\times 10^{19}$ cm$^{-3}$ at 2 K (6.25 $\times 10^{19}$ cm$^{-3}$ at 20 K),  which are compatible with the above estimate of $n$ at 12 K.

\subsection{C. Band structure}

\begin{figure}[ht]
\centering\includegraphics[width=1.0\columnwidth,keepaspectratio]{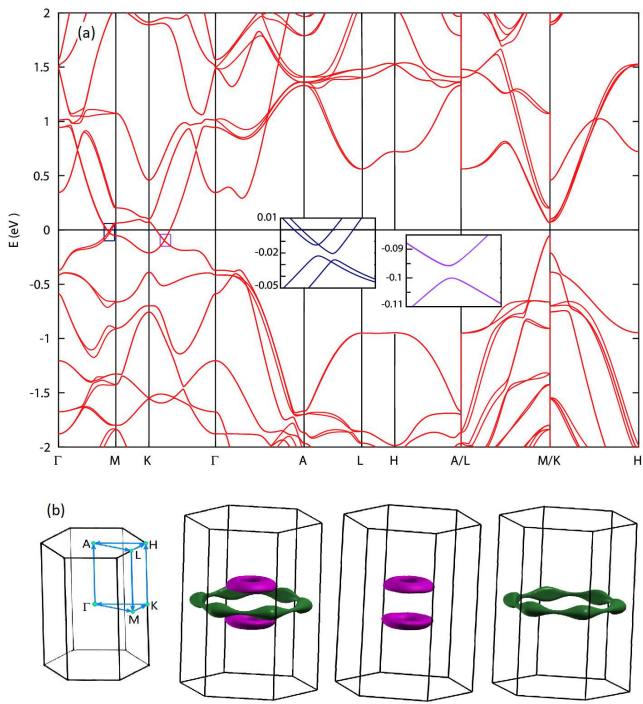}
\caption{(Color online) (a) Band structure (with SOC), and (b) Fermi pockets (without SOC) of YbPdAs obtained from DFT calculations, assuming well-localized 4\textit{f} electrons on the Yb$^{3+}$ ions. The Brillouin zone with labeled high-symmetry points is also shown in (b). \tcr{The insets of (a) highlight the two features near the Fermi level where the bands nearly cross. }  }
\label{fig9}
\end{figure}

To obtain further information about the electronic structure, we have performed DFT calculations for YbPdAs, where the 4\textit{f}-electrons are assumed to be well-localized (core electrons). 
The calculated band structure is shown in Fig.~\ref{fig9}(a) where SOC is taken into account. It can be seen that there are two doubly degenerate bands crossing the Fermi level, which are split into four by the SOC. At first glance, without SOC these two bands go through each other forming two band-crossings at around 22 meV (navy) and 100 meV (magenta) below $E\rm_F$, respectively, which further split to form eight band crossings under SOC. However, careful examination shows that all of them are actually separated even without SOC with small separations. With the inclusion of SOC, the separations are enhanced to 10.0 meV (navy) and 5.0 meV (magenta), as shown in the inset of Fig.~\ref{fig9}(a). 

 The Fermi pockets in the absence of SOC are shown in Fig.~\ref{fig9}(b) with two oblate hole pockets (magenta) and one circular electron pocket (olive). Each of them is further split by the SOC, but their shape is maintained. Note that the hole pockets are not visible in the band structure shown above, which only contains information along the high-symmetry path. The volume of the Fermi pockets is relatively small, which is consistent with the low carrier concentration, according to Luttinger's theorem \cite{Luttinger}. The carrier densities for electrons and holes are calculated to be 7.39 $\times$ 10$^{20}$ and 7.82 $\times$ 10$^{20}$ cm$^{-3}$, respectively, which are moderately small, comparable to those of semimetallic YbAs \cite{YbAs}.
 However, a small carrier density is not likely the origin of the increase of resistivity upon cooling at intermediate temperatures, since purely metallic behavior is found in LuPdAs, which has similarly small Fermi surfaces. This points to the importance of \textit{f}-electron effects, in particular Kondo scattering. 
The density of Yb$^{3+}$-ions in YbPdAs is approximately 1.81 $\times$ 10$^{22}$ cm$^{-3}$, which is about two orders of magnitude larger than the carrier densities obtained from calculations.

\section{IV. Summary}

We have synthesized polycrystalline sample of YbPdAs and characterized the physical properties. This compound crystallizes in the hexagonal ZrNiAl-type structure (space group \emph{P}$\bar{6}2$\emph{m}), in which the in plane Yb atoms form a distorted Kagome lattice with relatively large nearest neighbor Yb-Yb distances out of the hexagonal plane. 

Temperature dependent measurements of the specific heat, magnetic susceptibility, and resistivity show evidence for a second order AFM transition at $T_\mathrm{N}$ = 6.6 K, which is only slightly suppressed by applied magnetic fields up to 9 T. The Kondo effect is evidenced by a $-\ln T$ dependence of the resistivity at intermediate temperatures and reduced magnetic entropy at $T_\mathrm{N}$. Magnetic fluctuations above $T_\mathrm{N}$ = 6.6 K are indicated by a logarithmic increase of $C_{el}/T$ upon cooling below 15 K. Furthermore, the carrier concentration is relatively small, as revealed by transport measurements and DFT calculations. These results suggest that YbPdAs is a good candidate for examining the interplay of magnetic frustration and the Kondo effect in semimetallic systems. Measurements of the anisotropic properties on single crystals are highly desirable, as well as microscopic probes of the magnetic ground state and excitations using neutron scattering and muon-spin relaxation.

\section{V. Acknowledgments}
\begin{acknowledgments}
 This work was supported by the National Natural Science Foundation of
China (Grants No. 11974306, No. 12034017, and No. 12174332), the National Key R\&D
Program of China (Grant No. 2017YFA0303100), the Key
R\&D Program of Zhejiang Province, China (Grant No.
2021C01002),  and the Zhejiang Provincial Natural Science
Foundation of China (Grant No. R22A0410240). T.S. acknowledges support from the Natural Science Foundation of Shanghai (Grants No. 21ZR1420500 and No. 21JC1402300).
\end{acknowledgments}

\end{document}